\documentclass[prd,twocolumn,tightenlines,showpacs,eqsecnum]{revtex4}
\usepackage{amssymb}
\usepackage{amsmath}
\usepackage[dvips,final]{graphicx}
\usepackage{epsfig}
\usepackage{bm}

\begin{document}

\title{Pion distribution amplitude from holographic QCD
and the electromagnetic form factor $F_{\pi}(Q^2)$ \\
}
\author{\textbf{S.~S.~Agaev}}
\email{agaev_shahin@yahoo.com} \affiliation{Institute for Physical
Problems, Baku State University,
 Z. Khalilov St.\ 23, Az-1148
Baku, Azerbaijan}
\author{\textbf{M.~A.~Gomshi~Nobary}}
\email{mnobary@razi.ac.ir} \affiliation{Department of Physics,
Faculty of Science, Razi University, Kermanshah, Iran}

\begin{abstract}
The holographic QCD prediction for the pion distribution amplitude
(DA) $\varphi_{hol}(u)$ is used to compute the pion spacelike
electromagnetic form factor $F_{\pi}(Q^2)$ within the QCD
light-cone sum rule method. In calculations the pion's
renormalon-based model twist-4 DA, as well as the asymptotic
twist-4 DA are employed. Obtained theoretical predictions are
compared with experimental data and with results of the
holographic QCD.
\end{abstract}

\maketitle

\section{Introduction}

\label{sec:int}

Recently holographic dual models of QCD were proposed
\cite{Pol03,Erl05,Erl06,Rol05,Br04,Br06,Son06} and applied for
investigations to the hadronic physics. This approach is based on
the AdS/CFT correspondence \cite{Mal98}, and is aiming to
construct dual models of QCD in 5-dimensional Anti-de Sitter (AdS)
space. Invented models incorporate important features of QCD, as
confinement and chiral symmetry breaking, and allow one to
calculate numerous hadronic properties including masses, decay
constants, couplings of various mesons.

The holographic models of QCD were employed for investigation of
mesons electromagnetic form factors (FFs), as well. Calculations
of the pion and $\rho$-meson FFs in Refs.\
\cite{Rad06,Rad07,Rad071,Br07,Leb07} were carried out in the
framework of two models, namely in hard-wall and soft-wall ones
depending on the procedure to impose the infrared cutoff on the
fifth (holographic) dimension of the AdS space.

It is remarkable, that the holographic QCD (HQCD) predicts also
mesons' valence wave functions thus providing analytic
approximation to QCD \cite{Br07}. By this way, the HQCD prediction
for the pion distribution amplitude (DA) $\varphi_{hol}(u)$ was
derived. It differs from the pion asymptotic DA $\varphi_{asy}(u)$
found from the perturbative QCD (PQCD) evolution \cite{Br79}, and
due to the broader shape should increase the magnitude of the
leading twist QCD prediction for the pion electromagnetic FF.

In the present work the DA $\varphi_{hol}(u)$ is used to compute
the pion spacelike electromagnetic form factor in the context of
the QCD light-cone sum rule (LCSR) method with twist-6 accuracy by
taking into account $O(\alpha_{\rm{s}})$ order correction to the
leading twist term. The twist-4 DA of the pion, necessary to
determine the twist-4 contribution to the FF, is derived in the
context of the renormalon method. The twist-4 contribution is also
modeled employing the asymptotic form of the pion twist-4 DA
following from the conformal expansion.

The paper is structured as follows: In Sec.\ \ref{sec:LCSR}
general expressions for the form factor $F_{\pi}(Q^2)$ in LCSR
method are presented. Section \ref{sec:DA} is devoted to
calculation of the pion twist-4 DA employing the renormalon
approach. In Sec.\ \ref{sec:num} we present results of numerical
computations and compare them with the experimental data and
predictions of the holographic QCD. Section\ \ref{sec:conc}
contains our conclusions.

\section{ The pion electromagnetic form factor in the QCD LCSR method}

\label{sec:LCSR}

It is known that the QCD LCSR method is one of the powerful tools
to evaluate nonperturbative components of exclusive quantities
\cite{LCSR}. The LCSR expression for the pion electromagnetic FF
was derived in Refs.\ \cite{BH94,Br00,Bij}: with the twist-6
accuracy it has the following form
\begin{equation}  \label{eq:1}
F_\pi (Q^2)=F_\pi ^{(2)}(Q^2)+F_\pi ^{(2,\alpha
_{\rm{s}})}(Q^2)+F_\pi ^{(4)}(Q^2)+F_\pi ^{(6)}(Q^2),
\end{equation}
where $Q^2=-q^2$, $q$ being the four-momentum of the virtual
photon in the process $\gamma^*\pi^{\pm} \to \pi^{\pm}$. In Eq.\
(\ref{eq:1}) $F_{\pi}^{(n)}(Q^2)$ is the twist-$n$ contribution to
the FF; $F_\pi ^{(2,\alpha _{\rm{s}})}(Q^2)$ is
$O(\alpha_{\rm{s}})$ order correction to the twist-2 part.

The leading twist (twist-2) light-cone sum rule for $F_\pi (Q^2)$
at the leading order is given by the expression

\begin{equation}  \label{eq:2}
F_\pi ^{(2)}(Q^2)=\int_{u_0}^1du\varphi^{(2)}(u,\mu _F^2)\exp
\left[ -\frac{ \overline{u}Q^2}{uM^2}\right],
\end{equation}
where
\begin{equation}  \label{eq:3}
u_0=\frac{Q^2}{s_0+Q^2}.
\end{equation}
In Eqs. (\ref{eq:2}) and (\ref{eq:3}) $\varphi^{(2)}(u,\mu _F^2)$
is the pion leading twist DA; $s_0$ is the duality interval, $M^2$
is the Borel variable and $\overline{u} \equiv 1-u$.

The $O(\alpha_{\rm{s}})$ order correction to the leading twist
term was obtained in Ref.\ \cite{Br00}. Then the whole twist-2
contribution to the form factor
\[
{\mathbf F}_{\pi}^{(2)}(Q^2)=F_\pi ^{(2)}(Q^2)+F_\pi ^{(2,\alpha
_{\rm{s}})}(Q^2)
\]
can be written as the sum of the soft and hard parts
\[
{\mathbf F}_\pi ^{(2)}(Q^2)=\int_{0}^{1}du\varphi^{(2)}(u,\mu
_F^2)\left [ \Theta(u-u_0){\mathit
F}_{soft}^{(2)}(u,M^2,s_0)\right.
\]
\begin{equation}
\left. +\Theta(u_0-u){\mathit F}_{hard}^{(2)}(u,M^2,s_0)\right ].
\label{eq:3a}
\end{equation}
The soft part of the LCSR contains entirely the leading order
twist-2 contribution and some piece of the next-to-leading order
correction to the twist-2 term. Stated differently, the expression
presented in Eq.\ (\ref{eq:2}) is the pure soft contribution to
the FF. The twist-2 term (\ref{eq:3a}) is linear in the pion DA.
Its hard component at high momentum transfers $Q^2 \gg s_0$ leads
to the PQCD prediction for the FF quadratic in DA \cite{Br00}. For
example, choosing as $\varphi^{(2)}(u,\mu_F^{2})$ the pion
asymptotic DA $\varphi_{asy}(u)$, from the hard part of the LCSR
one regains the well-known asymptotic prediction of the PQCD
\cite{Br79,Rad80}
\[
F_{\pi}^{asy}(Q^2)=\frac{8\pi\alpha_{\rm s}(Q^2)f_{\pi}^2}{Q^2},
\]
with $f_{\pi}=0.132 \,\, {\rm GeV}$ being the pion decay
constant.

Further details of calculations and explicit expressions for
${\mathit F}_{soft}^{(2)}(u,M^2,s_0)$ and ${\mathit
F}_{hard}^{(2)}(u,M^2,s_0)$ can be found in Ref.\ \cite{Br00}.

The twist-4 term $F_\pi ^{(4)}(Q^2)$ is determined as \cite{Bij}
\[
F_\pi ^{(4)}(Q^2)=\int_{u_0}^1du\frac{\phi_4(u,\mu _F^2)}{%
uM^2}\exp \left[ -\frac{\overline{u}Q^2}{uM^2}\right]
\]
\begin{equation}  \label{eq:4}
+\frac{u_0 \phi_4(u_0,\mu _F^2)}{Q^2}e^{-s_0/M^2}.
\end{equation}
Here
\begin{equation}  \label{eq:5}
\phi_4(u,\mu _F^2)=2u\left[ \frac d{du}\varphi _2^{(4)}(u,\mu
_F^2)-\overline{u}\frac{d^2}{du^2}\varphi _2^{(4)}(u,\mu _F^2)
\right],
\end{equation}
and $\varphi_2^{(4)}(u,\mu _F^2)$ is the pion two-particle twist-4
DA.

The factorizable twist-6 contribution to the LCSR was computed in
Ref. \cite {Br00} in terms of the quark condensate density
\begin{equation}  \label{eq:6}
F_\pi ^{(6)}(Q^2)=\frac{4\pi \alpha _{\rm{s}}(\mu _R^2)C_F}{3f_\pi
^2Q^4} \left\langle 0\left| \overline{q}q\right| 0\right\rangle
^2,
\end{equation}
where $C_F=4/3$ is the color factor.

In the QCD LCSR method for the factorization and renormalization
scales the following values should be accepted:
\begin{equation}
\mu _{F}^{2}=\mu _{R}^{2}=\overline{u}Q^{2}+uM^{2}.  \label{eq:6a}
\end{equation}
Equations (\ref{eq:1})-(\ref{eq:6}) supplemented by the
prescription (\ref{eq:6a}) form a basis for investigation of the
pion electromagnetic FF in the QCD LCSR method.

\section{The pion distribution amplitudes}
\label{sec:DA}

The light-cone two-particle distribution amplitudes of the pion
are defined through the light-cone expansion of the matrix element
\begin{widetext}
\[
\left\langle 0\left| \overline{d}(x_{2})\gamma _{\nu }\gamma
_{5}\left[ x_{2},x_{1}\right] u(x_{1})\right| \pi
^{+}(p)\right\rangle=if_{\pi }p_{\nu
}\int_{0}^{1}due^{-iupx_{1}-i\overline{u}px_{2}}\left[ \varphi
^{(2)}(u,\,\mu _{F}^{2}) +\Delta ^{2}\varphi _{1}^{(4)}(u,\,\,\mu
_{F}^{2})+O(\Delta ^{4})\right]
\]
\begin{equation}
+if_{\pi }\left( \Delta _{\nu }(p\Delta )-p_{\nu }\Delta
^{2}\right) \int_{0}^{1}due^{-iupx_{1}-i\overline{u}px_{2}}
 \left[\varphi _{2}^{(4)}(u,\,\,\mu _{F}^{2})+O(\Delta
^{4})\right], \label{eq:7}
\end{equation}
\end{widetext}
where $\varphi _{1}^{(4)}(u,\,\,\mu _{F}^{2}),\,\,\varphi
_{2}^{(4)}(u,\,\,\mu _{F}^{2})$ are two-particle twist-4 DAs,
$\Delta =x_{1}-x_{2}$, and we use the notation $\left[
x_{2},x_{1}\right] $ for the Wilson line connecting the points
$x_{1}$ and $x_{2}$.

The standard method to handle meson DAs is modeling them employing
the conformal expansion. Then for the leading twist pion DA we get
\cite{Br79}
\begin{equation}
\varphi^{(2)}(u,\,\mu _{F}^{2})=\varphi _{asy}(u)\left[ 1+\sum_{n=2,4..}^{%
\infty }b_{n}(\mu _{F}^{2})C_{n}^{3/2}(u-\overline{u})\right].
\label{eq:8}
\end{equation}
Here $\varphi _{asy}(u)$ is the PQCD asymptotic DA of the pion
\[
\varphi _{asy}(u)=6u\overline{u},
\]
and $C_{n}^{3/2}(\xi )$ are the Gegenbauer polynomials. The
functions $b_{n}(\mu _{F}^{2})$ determine the evolution of
$\varphi^{(2)}(u,\,\mu _{F}^{2})$ on the factorization scale $\mu
_{F}^{2}$ \cite{Br79},
\[
b_{n}(\mu _{F}^{2})=b_n^{0}\left[ \frac{\alpha _{\rm{s}}(\mu
_{F}^{2})}{\alpha _{\rm{s}}(\mu _{0}^{2})}\right] ^{\gamma
_{n}/\beta _{0}},
\]
\begin{equation}
\gamma _{n}=C_{F}\left[ 1-\frac{2}{(n+1)(n+2)}%
+4\sum_{j=2}^{n+1}\frac{1}{j}\right] .  \label{eq:8a}
\end{equation}
In the above $\gamma _{n}$ are the anomalous dimensions, $\mu
_{0}^{2}$ is the normalization scale, and $b_n^{0} \equiv
b_{n}(\mu _{0}^{2})$.  The expansion over the conformal spin can
also be performed for the higher twist DAs (see, for example,
Refs. \cite {BB99,Ball99}).

 An alternative way to find the higher twist DAs is the
renormalon approach \cite{An00,BGG}. The renormalon approach
employs the assumption that the infrared renormalons in the
leading twist coefficient functions should cancel the ultraviolet
renormalons in the matrix elements of twist-4 operators in a
relevant operator product expansion. Such cancellation was proved
by explicit calculations in the case of the simple exclusive
amplitude involving pseudoscalar and vector mesons \cite{BGG}. It
turned out that this is enough to find the full set of two- and
three-particle twist-4 DAs of pseudoscalar and vector mesons in
terms of their leading twist DAs.  Higher twist DAs of some of the
mesons were computed using the renormalon technique in the papers
\cite{Ag051,BB1,BB2}.

In the renormalon-based model the pion twist-4 DAs are given by
the formulas \cite{BGG}
\begin{widetext}
\begin{equation}
\varphi _{1}^{(4)}(u,\mu _{F}^{2})=\frac{\delta ^{2}}{6}\int_{0}^{1}dv%
\varphi ^{(2)}(v,\mu _{F}^{2})\left\{ \frac{1}{v^{2}}\left[
u+(v-u)\ln \left( 1-\frac{u}{v}\right) \right] \theta (v>u)
+\frac{1}{\overline{v}^{2}}\left[ \overline{u}+(u-v)\ln \left( 1-%
\frac{\overline{u}}{\overline{v}}\right) \right] \theta
(v<u)\right\}, \label{eq:9}
\end{equation}
\begin{equation}
\varphi _{2}^{(4)}(u,\mu _{F}^{2})=-\frac{\delta ^{2}}{6}\int_{0}^{1}dv%
\varphi ^{(2)}(v,\mu _{F}^{2})\left[ \left( \frac{u}{v}\right)
^{2}\theta
(v>u)+\left( \frac{\overline{u}}{\overline{v}}\right) ^{2}\theta (v<u)%
\right]. \label{eq:10}
\end{equation}
\end{widetext}

As is seen the renormalon model for the twist-4 DAs depends only
on one free parameter $\delta ^{2}$. It is related to the matrix
element of the local operator
\[
\left\langle 0\left| \overline{d}\gamma _{\nu
}ig\widetilde{G}_{\mu \rho }u\right| \pi ^{+}(p)\right\rangle
=\frac{1}{3}f_{\pi }\delta ^{2}\left[ p_{\rho }g_{\mu \nu }-p_{\mu
}g_{\rho \nu }\right],
\]
\begin{equation}
\delta^2(\mu_F^2)=\delta^2(\mu_0^2) \left[ \alpha_{\rm
s}(\mu_F^2)/ \alpha_{\rm s}(\mu_0^2)\right]^{8C_F/3\beta_0}
\label{eq:10a},
\end{equation}
and at $\mu_0^2=1 \,\, {\rm GeV}^2$ was estimated from various
2-point QCD sum rules \cite{BB1,NS94}
\begin{equation}
\delta ^{2}(\mu _{0}^{2}) \equiv \delta_{0}^{2}=0.18 \pm
0.06\,\,\rm{GeV}^{2}.\label{eq:10b}
\end{equation}
In Eq.\ (\ref{eq:10b}) we use the latest available estimation for
$\delta_{0}^{2}$ \cite{BB1}. In what follows we also do not show
explicitly a dependence of the parameter $\delta^2$ on the
factorization scale $\mu_F^2$.

In the case of the pion leading twist DA with two non-asymptotic
terms ($b_2(\mu_F^2),\, b_4(\mu_F^2) \neq 0$, and $b_n=0, n>4$)
these higher twist distributions were calculated in Ref.\
\cite{Ag051} (see Erratum in Ref.\ \cite{Ag052}). Let us rewrite
the twist-4 DA $\varphi_2^{(4)}(u, \mu_{F}^2)$ obtained in Ref.\
\cite{Ag051}, and solely relevant for our present studies, in the
compact form
\begin{widetext}
\[
\varphi _{2}^{(4)}(u,\mu _{F}^{2})=\delta ^{2}\left[
u\overline{u}+u^{2}\ln u+\overline{u}^{2}\ln \overline{u} \right.
\left. +6b_{2}(\mu _{F}^{2})\left( u^{2}\ln u+
\overline{u}^{2}\ln \overline{u}+u\overline{u}+\frac{5}{3}u^{2}\overline{u}%
^{2}\right) \right.
\]
\begin{equation}
\left. +3b_{4}(\mu _{F}^{2})\left( 5u^{2}\ln
u+5\overline{u}^{2}\ln
\overline{u}+5u\overline{u}+\frac{77}{6}u^{2}\overline{u}^{2}-21u^{3}\overline{u}%
^{3}\right) \right]. \label{eq:11}
\end{equation}
Then Eq.\ (\ref{eq:5}) is given by the expression:
\[
\phi_{4}(u,\mu _{F}^{2})=2\delta ^{2}\left\{ -4u\overline{u}%
-4u\overline{u}\ln\overline{u}+2u(2u-1)\ln u \right.
\left.+4b_{2}(\mu _{F}^{2})\left[ -11u\overline{u}%
+35u^{2}\overline{u}-40u^{3}\overline{u}-6u\overline{u}\ln \overline{u}%
+3u(2u-1)\ln u\right] \right.
\]%
\begin{equation}
\left. +b_{4}(\mu _{F}^{2})\left[ -137u\overline{u}+917u^{2}\overline{u}%
-3073u^{3}\overline{u}\right.
\left. +4347u^{4}\overline{u}-2268u^{5}\overline{u}-60u%
\overline{u}\ln \overline{u}+30u(2u-1)\ln u\right] \right\}.
\label{eq:12}
\end{equation}
\end{widetext}
\begin{figure}[tbp]
%
\centering\epsfig{file=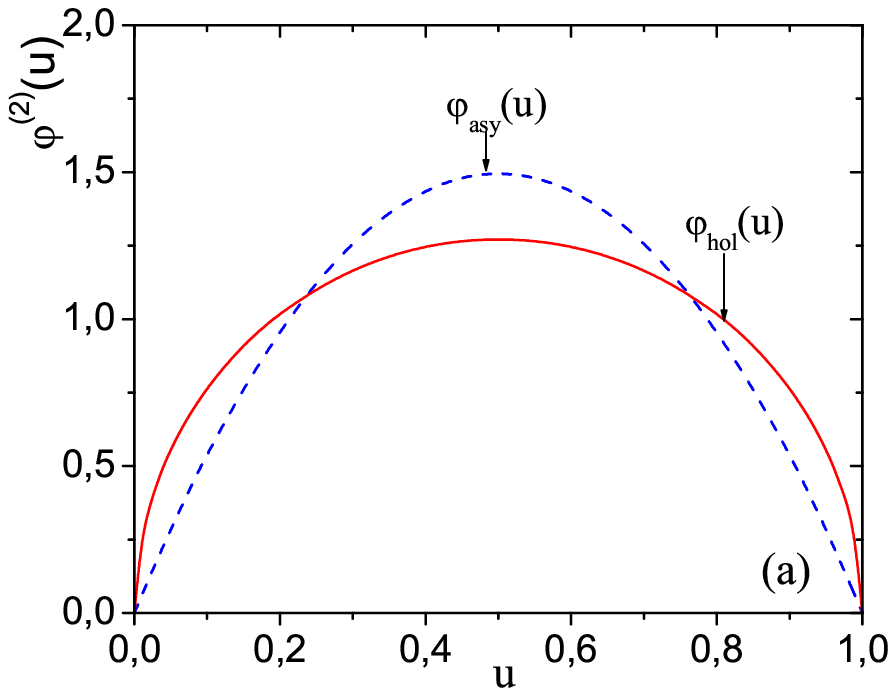,height=6cm,width=8.0cm,clip=}
\hspace{0.cm}
\centering\epsfig{file=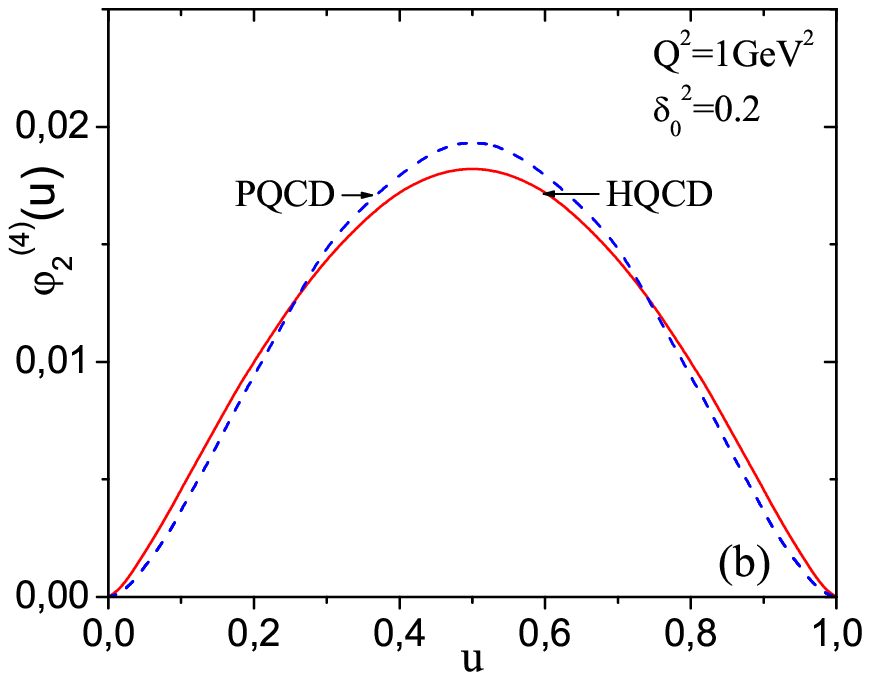,height=6cm,width=8.0cm,clip=}
\vspace{ -0.6cm} \caption{{\bf (a)} The leading twist
distributions of the pion, {\bf (b)} the twist-4 DAs obtained from
$\varphi_{asy}(u)$ (the dashed curve) and $\varphi_{hol}(u)$ (the
solid curve) by means of the renormalon approach.} \label{fig:tw2}
\end{figure}

\begin{figure}[tbp]
%
\centering\epsfig{file=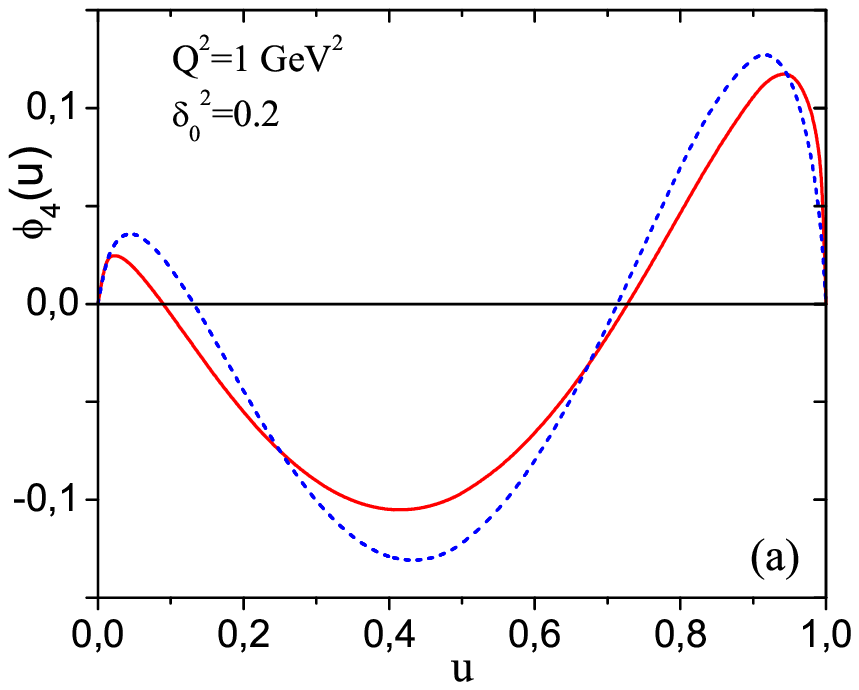,height=6cm,width=8.0cm,clip=}
\hspace{0.cm}
\centering\epsfig{file=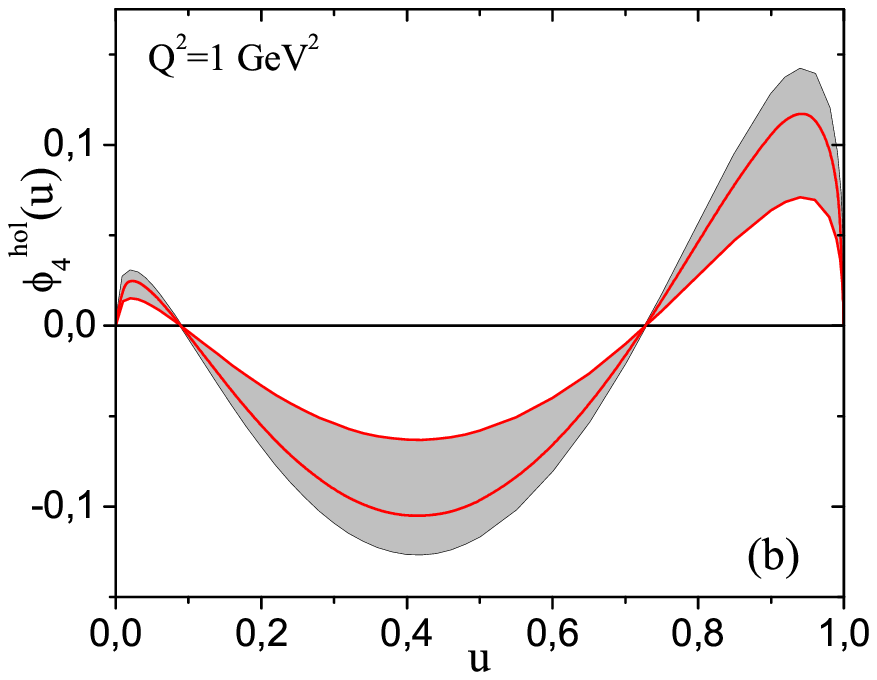,height=6cm,width=8.0cm,clip=}
\vspace{ -0.6cm} \caption{{\bf (a)} The twist-4 function
$\phi_4(u)$ obtained in the renormalon-based model using the PQCD
asymptotic DA (the dashed line) and the distribution amplitude of
the HQCD (the solid line); {\bf (b)} the dependence of the twist-4
function $\phi_4^{hol}(u)$ on the input parameter $\delta_0^2$.}
\label{fig:twist4}
\end{figure}

The distribution amplitude of the pion obtained within the HQCD
\begin{equation}
\varphi_{hol}(u)=\frac{8}{\pi}\sqrt{u\overline{u}}, \label{eq:12a}
\end{equation}
in the renormalon approach leads to the following twist-4 DA
\[
\varphi _{2}^{(4)}(u)=-\frac{8 \delta ^{2}}{3\pi }\left[
u^{2}\left( \sqrt{ \frac{\overline{u}}{u}}-\arctan
\sqrt{\frac{\overline{u}}{u}}\right) \right.
\]
\begin{equation}
\left. +\overline{u}^{2}\left( \sqrt{\frac{u}{\overline{u}}}-\arctan \sqrt{\frac{u}{%
\overline{u}}}\right) \right]. \label{eq:13}
\end{equation}
As a result, for the twist-4 function $\phi_4^{hol}(u)$, we get
\[
\phi_{4}^{hol}(u)=\frac{4\delta ^{2}}{3\pi }\left[u(1+8u)\sqrt{\frac{%
\overline{u}}{u}}+
(3-11u+8u^{2})\sqrt{\frac{u}{\overline{u}}}\right.
\]
\begin{equation}
\left. +8u(2u-1)\arctan \sqrt{\frac{\overline{u}}{u}}-16u\overline{u}%
\arctan \sqrt{\frac{u}{\overline{u}}}\right]. \label{eq:14}
\end{equation}

\begin{figure}[tbp]
%
\centering\epsfig{file=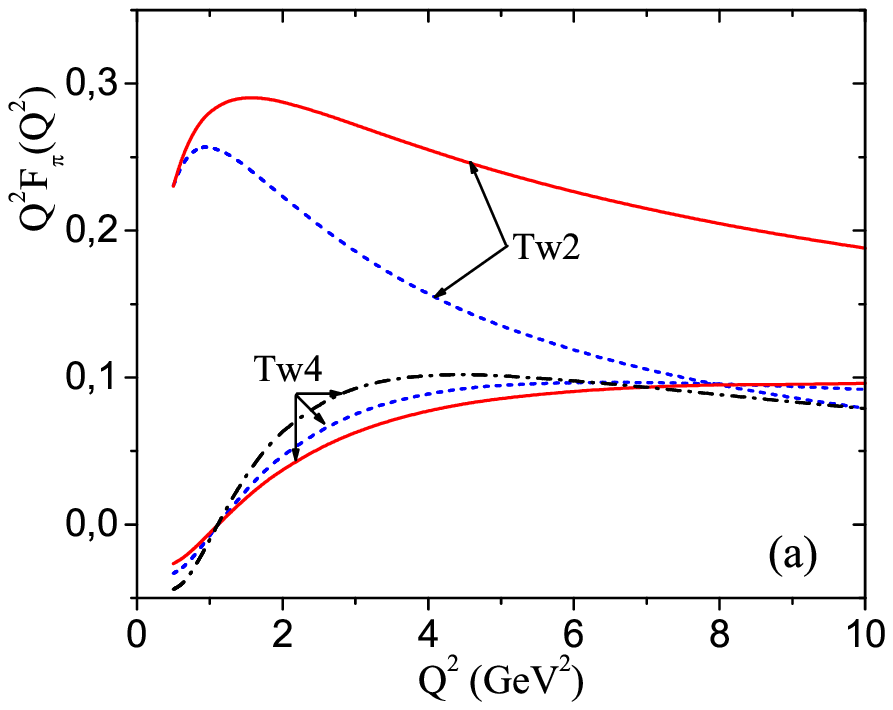,height=6cm,width=8.0cm,clip=}
\hspace{0.cm}
\centering\epsfig{file=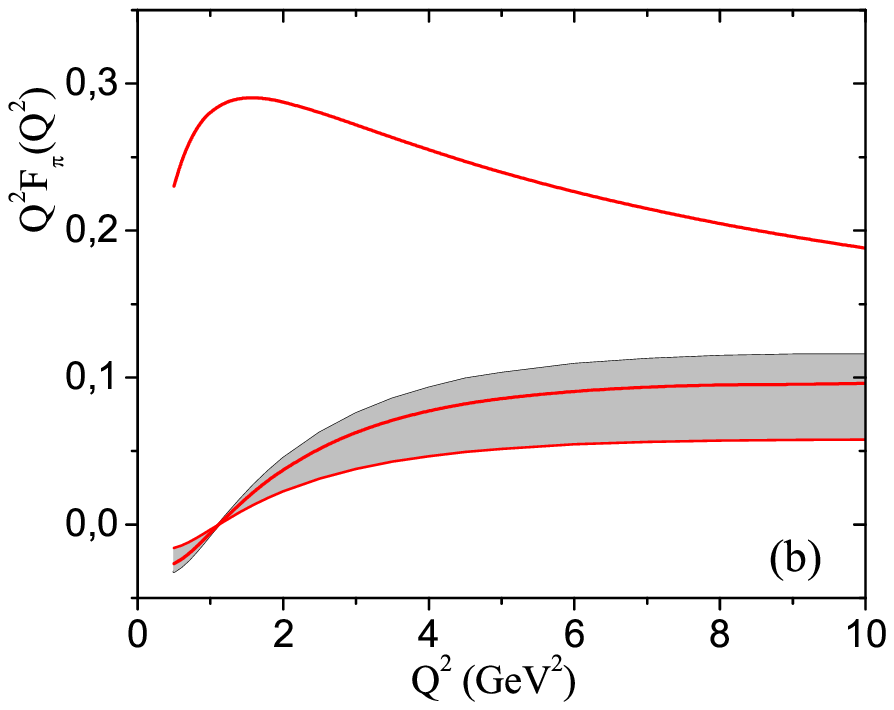,height=6cm,width=8.0cm,clip=}
\vspace{ -0.6cm} \caption{{\bf (a)} The leading and twist-4
contributions to the scaled pion form factor. In the calculations
the PQCD DAs (the dashed lines) and DAs of the HQCD (the solid
lines) are used. For comparison the asymptotic twist-4 term (the
dot-dashed line) is also shown. All twist-4 terms are obtained
using $\delta_{0}^{2}=0.2$. {\bf (b)} The leading and twist-4
contributions to the FF found employing the functions
$\varphi_{hol}(u)$ and $\phi_4^{hol}(u)$. The shaded area
demonstrates the allowed values of the twist-4 term obtained by
varying the parameter $\delta_0^2$ within the limits
(\ref{eq:10b}).} \label{fig:fig3}
\end{figure}

The leading twist distributions $\varphi_{asy}(u)$ and
$\varphi_{hol}(u)$ and the corresponding twist-4 DAs
$\varphi_2^{(4)}(u)$ are depicted in Fig.\ \ref{fig:tw2}. We see
that as compared with $\varphi_{asy}(u)$ the holographic DA
$\varphi_{hol}(u)$ is enhanced in the end-point domains. The
difference between the twist-4 DAs shown in Fig.\ \ref{fig:tw2}(b)
is mild. As a result, the twist-4 functions $\phi_4(u)$ and
$\phi_4^{hol}(u)$ that determine the twist-4 contribution to the
form factor are close to each other (Fig.\ \ref{fig:twist4}(a)).
Nevertheless, we should note that the function $\phi_4^{hol}(u)$
is enhanced in the end-point region $u \to 1$. In Fig.\
\ref{fig:twist4}(b) we demonstrate the variation of the form of
$\phi_{4}^{hol}(u)$ depending on the chosen value of the parameter
$\delta_{0}^{2}$.

The twist-4 DA of the pion can be modeled in the context of the
conformal expansion as well. The lowest conformal spin, i.e. the
asymptotic form of the twist-4 DA $\varphi_2^{(4)}(u)$ was
employed in Ref.\ \cite{Bij} for calculation of the twist-4
function (\ref{eq:5}) with the following result

\begin{equation}
\phi_{4}^{asy}(u)=\frac{20}{3}\delta ^{2}u\overline{u}\left[
1-u(7-8u)\right]. \label{eq:15}
\end{equation}
We use the function $\phi_4^{asy}(u)$ for calculation of the
asymptotic twist-4 term.

\section{Numerical results}
\label{sec:num}

In order to perform numerical computations we should fix various
parameters appearing in the relevant expressions. Namely, we take
the Borel parameter equal to $M^2=1\, {\rm GeV^2}$ and accept for
the factorization and renormalization scales the values presented
in Eq.\ (\ref{eq:6a}). For the QCD coupling $\alpha \rm{_{s}}(\mu
_{R}^{2})$ the two-loop expression with $\Lambda
_{3}=0.34\,\,\rm{GeV}$ is used. The value of the duality parameter
$s_{0}=0.7\,\,\rm{GeV}^{2}$ is borrowed from the QCD sum rule for
the correlator of two $\overline{u}\gamma _{\mu }\gamma _{5}d$
currents \cite{SVZ}. The normalization scale is set equal to $\mu
_{0}^{2}=1\;\rm{GeV}^{2}$.

\begin{figure}[tbp]
%
\centering\epsfig{file=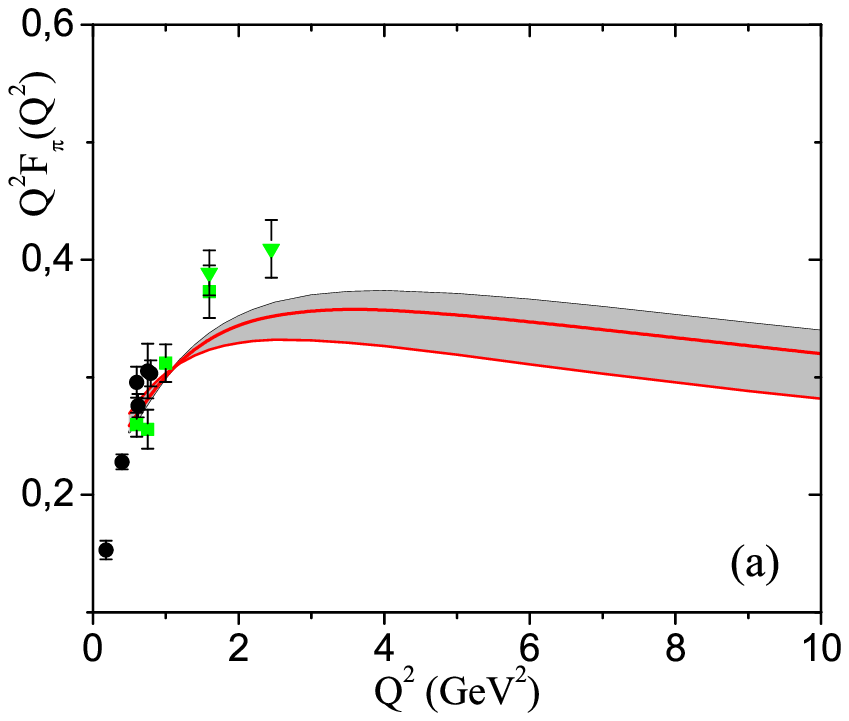,height=6cm,width=8.0cm,clip=}
\hspace{0.cm}
\centering\epsfig{file=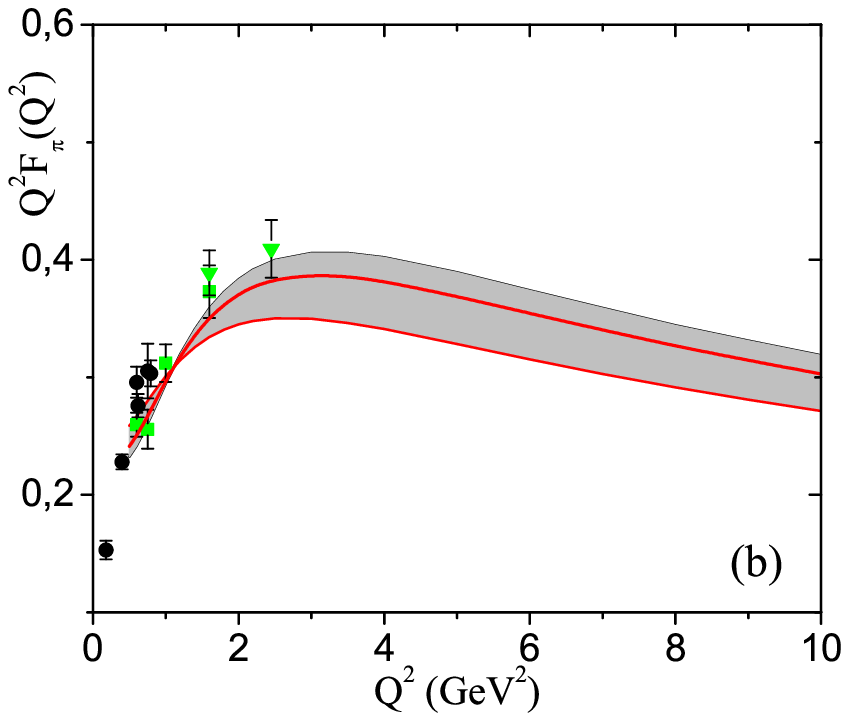,height=6cm,width=8.0cm,clip=}
\vspace{ -0.6cm} \caption{{\bf (a)} The pion scaled FF as a
function of $Q^2$. It is computed using the HQCD distribution
amplitude $\varphi_{hol}(u)$ and the renormalon-based twist-4
function $\phi_4^{hol}(u)$. The shaded area shows allowed limits
of the form factor. For the central solid line $\delta_0^{2}=0.2$.
{\bf (b)} The pion FF obtained using the asymptotic twist-4 term
and the holographic ones for the remaining contributions from Eq.\
(\ref{eq:1}). For the central solid line $\delta_0^{2}=0.2$.
Concerning the data - see discussion in the text .}
\label{fig:fig4}
\end{figure}

\begin{figure}[tbp]
\epsfxsize=8 cm \epsfysize=7 cm \centerline{\epsffile{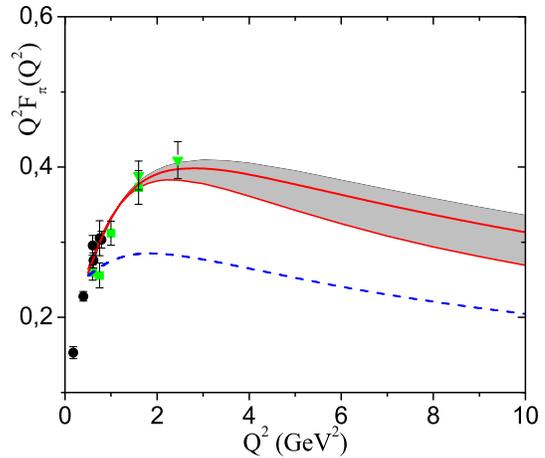}}
\vspace{-0.5cm} \caption{The pion scaled FF as a function of
$Q^2$. The central solid line is the LCSR prediction obtained
employing the PQCD leading twist DA with $b_2^0=0.25,\,\,b_4^0=0$
and corresponding renormalon-based twist-4 function
$\phi_4(u,Q^2)$. The shaded area is obtained by varying
$\delta_0^{2}$ within the allowed limits at fixed
$b_2^0=0.25,\,\,b_4^0=0$; for the central line $\delta_0^{2}=0.2$.
For comparison the LCSR result found using $\varphi_{asy}(u)$ is
also shown (the dashed line, $\delta_0^{2}=0.2$).}
\label{fig:fig5}
\end{figure}

\begin{figure}[tbp]
\epsfxsize=8 cm \epsfysize=7 cm \centerline{\epsffile{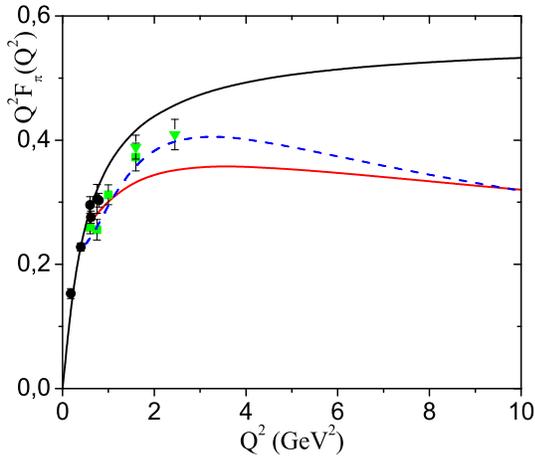}}
\vspace{-0.5cm} \caption{The pion scaled FF as a function of
$Q^2$. The lower solid line is prediction of the LCSR method with
the distribution  amplitude of the HQCD. The dashed line is the
same the HQCD result, but the HQCD twist-4 term replaced by the
asymptotic twist-4 one. The upper solid line is the soft-wall
holographic QCD prediction from Ref.\ \cite{Br07}.}
\label{fig:fig6}
\end{figure}

Results of numerical computations are depicted in Figs.\
\ref{fig:fig3}-\ref{fig:fig6}. As is seen (Fig.\
\ref{fig:fig3}(a)) the leading twist LCSR contribution found
employing $\varphi_{hol}(u)$, due to the broader shape of the HQCD
distribution amplitude, is larger than the prediction obtained by
means of $\varphi_{asy}(u)$. The twist-4 terms computed  using the
twist-4 function $\phi_4^{hol}(u)$ and the asymptotic version of
Eq.\ (\ref{eq:12}) do not differ considerably from each other: one
observes some deviation of the solid curve from the result of the
standard QCD, i.e. from the dashed line. Here, the features of the
different twist-4 terms that have interesting consequences should
be emphasized. Thus, the asymptotic twist-4 term (the dot-dashed
line) in the region of low momentum transfers demonstrates more
rapid increasing on $Q^2$ than other twist-4 terms: in this region
among the twist-4 contributions the HQCD prediction is a flattest
and lowest one. Contrary, for the momentum transfers $Q^2 > 8\;
{\rm GeV}^2$ the HQCD twist-4 term overtakes other twist-4
contributions. What is important, in the whole region of
considering momentum transfers only the HQCD twist-4 term is
monotonically increasing function of $Q^2$.
All twist-4 terms are sensitive to the choice of the parameter
$\delta_{0}^{2}$. In Fig.\ \ref{fig:fig3}(b), as an example, we
plot the twist-2 and -4 terms generated by the HQCD distributions
and by shaded area show sensitivity of the twist-4 term to
$\delta_{0}^{2}$.

The pion scaled electromagnetic FF $Q^2F_{\pi}(Q^2)$ computed
employing the HQCD distribution amplitude and twist-4 function
$\phi_4^{hol}(u)$ is depicted in Fig.\ \ref{fig:fig4}(a). It is
clear that the enhancement in the leading twist term is not
sufficient to describe the existing data on $F_{\pi}(Q^2)$; the
corresponding LCSR curve runs below the data points. If we replace
the holographic twist-4 term by its asymptotic counterpart keeping
the remaining ones unchanged, then the final expression describes
the experimental results. Nevertheless such agreement can be
achieved only for the large values of the parameter
$\delta_{0}^{2}>0.2$. From this analysis it is legitimate to
conclude that the enhancement generated by the shape of the
function $\varphi_{hol}(u)$ in the relevant terms of Eq.\
(\ref{eq:1}) and the asymptotic twist-4 term is enough to explain
the data on $F_{\pi}(Q^2)$ in the LCSR treatment.

The pion electromagnetic form factor was analyzed within the LCSR
method in Ref.\ \cite{Ag051}, where in calculations model DAs with
two nonasymptotic terms and renormalon inspired twist-4 DAs were
used, and from comparison with the data, constraints on the input
parameters $b_2^0$ and $b_4^0$ were extracted. In the fitting
procedure in Ref. \cite{Ag051} both the old experimental results
\cite{Beb} and new ones from the Jefferson Lab. $F_{\pi}$
Collaboration \cite{Vol} were used. It is worth noting that the
pion form factor is not measured  directly and real measurements
of the process $\gamma^{*}p \to \pi^{+}n$ require an extrapolation
of the off-shell pion to the mass-shell. Because this
extrapolation becomes increasingly problematic as $Q^2$ increases
and at high momentum transfers the old results suffer from large
uncertainties, in the present work we remove them from
consideration. We also correct the data from the $F_{\pi}$
Collaboration reported in Ref.\ \cite{Tad}, and add two new points
extracted by $F_{\pi2}$ Collaboration \cite{Horn}: in Figs.\
\ref{fig:fig4}, \ref{fig:fig5} and \ref{fig:fig6} the rectangle
and triangle symbols denote the $F_{\pi}$ and $F_{\pi2}$ data,
respectively. Strictly speaking, the LCSR method is applicable in
the domain of momentum transfers $Q^2 \geq 1 \,\, {\rm GeV}^2$.
Nevertheless, we extend our numerical calculations to the region
$Q^2 < 1 \,\, {\rm GeV}^2$ to reveal properties of the leading and
twist-4 terms at low momentum transfers. The data below this
border (the solid circles) are included into the figures for
illustrative purposes and are irrelevant for our present
discussions.

We have repeated calculations of Ref.\ \cite{Ag051} using the pion
model DAs with two nonasymptotic terms. One of these FFs, that
corresponds to values of the input parameters
$b_2^0=0.25,\,\,b_4^0=0$ and $\delta_{0}^{2}=0.2$, is plotted in
Fig.\ \ref{fig:fig5}. The achieved agreement with new data of the
JLAB experiment is encouraging. It is easy to see that excluding
of the unprecise data from analysis and adding new ones results in
a shift of the parameter $b_2^0$ towards larger values than in
Ref.\ \cite{Ag051}. The similar curves can be obtained in the more
general case $b_2^0\neq 0, \,\,b_4^0\neq 0$ as well.

In the framework of the holographic QCD the pion electromagnetic
FF has been calculated in Refs.\ \cite{Br07,Rad071,Leb07}. The
soft-wall model expression for the form factor is especially
simple, and is given by the monopole form \cite{Br07}
\begin{equation}
F_{\pi}(Q^2)=\frac{4k^2}{4k^2+Q^2}, \label{eq:16}
\end{equation}
where $k=0.375 \, {\rm GeV}$.

In Fig.\ \ref{fig:fig6} we compare the LCSR predictions found
using the functions $\varphi_{hol}(u)$ and $\phi_{4}^{hol}(u)$
(the lower solid line) and $\varphi_{hol}(u)$ and
$\phi_{4}^{asy}(u)$ (the dashed line) with the prediction of the
soft-wall HQCD (the upper solid line). Let us note that the dashed
line has been obtained choosing the maximal allowed value of
$\delta_{0}^{2}=0.24$.

The first LCSR result with the HQCD distribution amplitude and
renormalon inspired twist-4 function in the range of the momentum
transfers $2 \leq Q^2 \leq 10\ {\rm GeV}^2 $ demonstrates the
scaling in accordance with $Q^2F_{\pi}(Q^2) \simeq {\rm const}$
law with a weak sign of the logarithmic running of the QCD
coupling. This feature can be explained as a consequence of the
dependence of the corresponding twist-4 term on $Q^2$ due to the
enhancement of the $\varphi_4^{hol}(u)$ at end-point regions. The
second LCSR prediction computed also in this work differs from the
first one only in the choice of the twist-4 term. It is evident
that the dashed line shows more rapid fall-off with $Q^2$ than the
lower solid line, though due to the choice $\delta_{0}^{2}=0.24$
is in agreement with the data. Such rapid falling is typical for
other LCSR curves obtained using the PQCD DAs (Fig.\
\ref{fig:fig5}).

In the LCSR method the soft component of the pion FF is modeled by
the leading order twist-2 term and some piece of the
$O(\alpha_{\rm s})$ order correction to the twist-2 term. The
leading order higher twist terms are soft contributions to the
pion FF as well. There are various approaches to model the soft
component of the FF. For example, in Ref.\ \cite{Agpion} the pion
FF was evaluated applying the standard hard-scattering approach
and the QCD running coupling method. It was demonstrated that the
Borel resummed expression of the form factor contains both the
soft and hard components, and in the asymptotic limit $Q^2 \to
\infty$ the standard PQCD result can be recovered. Alternatively,
in Ref.\ \cite{choi} soft and hard contributions to the pion
electromagnetic FF were calculated within the light-front quark
model and comparison with the AdS/CFT prediction was performed.
The authors reported on excellent agreement of their result with
the HQCD prediction.

Comparison of our results with the soft-wall holographic QCD
prediction reveals the discrepancy between them. First of all, the
prediction of the holographic QCD overestimates the JLAB data: a
situation with a hard-wall result is even worse \cite{Br07}.
Second, the HQCD prediction obeys the scaling $Q^{n}F_{\pi}(Q^2)
\to {\rm const}$ with $n \neq 2$. From our point of view,
emphasized also in Ref.\ \cite{choi}, for a reliable analysis one
should evolve the pion DA to a scale of the considering process,
which in our case is equal to a few ${\rm GeV}^2$. This procedure
inevitably demands to include nonasymptotic terms into the
holographic DA of the pion. In other words, for credible
phenomenological applications the holographic QCD version of Eq.\
(\ref{eq:8}), where the function $\varphi_{hol}(u)$ is only the
first term in a relevant expansion, is required. Higher order
corrections neglected in present investigations of the AdS/CFT
correspondence may improve a situation, too \cite{choi}.

For successful applications of the pion HQCD distribution
amplitude within the framework of traditional methods of the
perturbative QCD, one should prove that this DA describes also
other exclusive and semi-inclusive processes involving the pion.
The simplest such process is the pion electromagnetic transition
form factor $F_{\pi \gamma}(Q^2)$. In the context of the LCSR
method it was computed in Refs. \cite{Kh,Sch}: the result was
derived with twist-4 accuracy including the next-to-leading order
correction to the twist-2 term. In the context of the LCSR and
renormalon methods this form factor was analyzed in Refs.\
\cite{Ag052,Bak}. Preliminary calculations demonstrate that the
twist-2 term in the LCSR expression for $F_{\pi \gamma}(Q^2)$
found using $\varphi_{hol}(u)$ exceeds the measured experimental
data \cite{CLEO}. But contributions of the next-to-leading order
and twist-4 terms to the LCSR are negative and reduce the
magnitude of the leading twist contribution (see, for example,
\cite{Ag052}): hence an agreement with the data may be achieved. A
complete analysis of $F_{\pi \gamma}(Q^2)$ requires to compute
using $\varphi_{hol}(u)$ the renormalon-based twist-4 function
$\Phi^{(4)}(u,Q^2)$ \cite{Kh,Ag052} that determines the twist-4
contribution to the transition FF. This problem requires separate
investigations.

\section{Conclusions}
\label{sec:conc}

In the present work we have calculated the pion spacelike
electromagnetic form factor $F_{\pi}(Q^2)$ in the QCD LCSR
framework. We have used the distribution amplitude of the pion
derived in the holographic QCD, which has a broader shape than the
PQCD asymptotic DA. The twist-4 DA has been obtained applying the
renormalon inspired model of Ref.\ \cite{BGG}. The prediction of
the LCSR method obtained with the twist-6 accuracy by means of the
functions $\varphi_{hol}(u)$ and $\phi_{4}^{hol}(u)$ lies below
the data, and does not describe the experimental situation. The
agreement with the data can be achieved provided the HQCD twist-4
term is replaced by the asymptotic one and $\delta_{0}^{2}>0.2$ is
chosen.

We have also computed $F_{\pi}(Q^2)$ utilizing the pion PQCD
leading twist DAs with two nonasymptotic terms and corresponding
renormalon inspired twist-4 DAs. It has been proved that in this
case in the region $Q^2 > 1\,\, {\rm GeV}^2$ the QCD LCSR method
explains the new experimental situation emerged due to the data of
the JLAB experiment.

Comparison with results of the holographic QCD
\cite{Br07,Rad071,Leb07} has revealed an interesting tendency: the
holographic QCD predictions for $Q^2F_{\pi}(Q^2)$ obtained within
the both hard- and soft-wall models are increasing functions of
$Q^2$, whereas in the LCSR treatment at large momentum transfers
one finds an opposite picture. It is worth noting that the scaling
of the form factor calculated employing the functions
$\varphi_{hol}(u)$ and $\varphi_{4}^{hol}(u)$ is very promising.
Unfortunately, the corresponding curve runs below the data. In
order to enhance the magnitude of $F_{\pi}(Q^2)$ and reach an
agreement with the data additional contributions are needed. It
seems to us that such contributions may appear due to the
holographic QCD counterparts of the nonasymptotic terms in the
pion PQCD leading twist DA (\ref{eq:8}). If extracted, these terms
may improve scaling properties and normalization of the
holographic QCD predictions themselves.


\begin{thebibliography}{99}
\bibitem{Pol03} J.~Polchinski and M.~J.~Strassler,
Phys.\ Rev.\ Lett.\ {\bf 88}, 031601 (2002);
J. High Energy Phys. 05 (2003) 012.

\bibitem{Erl05} J.~Erlich, E.~Katz, D.~T.~Son and M.~A.~Stephanov, Phys.\
Rev.\ Lett.\ {\bf 95}, 261602 (2005).

\bibitem{Erl06} J.~Erlich, G.~D.~Kribs and I.~Low, Phys.\ Rev.\
D {\bf 73}, 096001 (2006).

\bibitem{Rol05} L.~Da Rold and A.~Pomarol, Nucl.\ Phys.\ {\bf B721}, 79
(2005);\  
 J. High Energy Phys. 01 (2006) 157.

\bibitem{Br04} S.~J.~Brodsky and G.~F.~de Teramond, Phys.\ Lett.\
{\bf B582}, 211 (2004); 
G.~F.~de Teramond and S.~J.~Brodsky, Phys.\ Rev.\ Lett.\ {\bf 94},
201601 (2005).

\bibitem{Br06} S.~J.~Brodsky and G.~F.~de Teramond,
Phys.\ Rev.\ Lett.\ {\bf 96}, 201601 (2006).


\bibitem{Son06} A.~Karch, E.~Katz, D.~T.~Son and M.~A.~Stephanov,
Phys.\ Rev.\ D {\bf 74}, 015005 (2006).

\bibitem{Mal98} J.~M.~Maldacena, Adv.\ Theor.\ Math.\ Phys.\ {\bf 2},
231 (1998).

\bibitem{Rad06} A.~V.~Radyushkin, Phys.\ Lett.\ {\bf B642}, 459 (2006).

\bibitem{Rad07} H.~R.~Grigoryan and A.~V.~Radyushkin, Phys.\ Lett.\ {\bf B650}, 421
(2007); 
Phys.\ Rev. D {\bf 76}, 095007 (2007).

\bibitem{Rad071} H.~R.~Grigoryan and A.~V.~Radyushkin, Phys.\ Rev. D {\bf 76}, 115007 (2007).

\bibitem{Br07} S.~J.~Brodsky and G.~F.~de Teramond, Phys.\ Rev.\ D {\bf 77}, 056007 (2008).

\bibitem{Leb07} H.~J.~Kwee and R.~L.~Lebed, J. High Energy Phys. 01 (2008) 027;
arXiv:0712.1811.

\bibitem{Br79} G.~P.~Lepage and S.~J.~Brodsky, Phys.\ Lett.\ {\bf B87},
359 (1979).

\bibitem{LCSR} I.~I.~Balitsky, V.~M.~Braun, and
A.~V.~Kolesnichenko, Nucl.\ Phys.\ {\bf B312}, 509 (1989);
V.~M.~Braun and I.~E.~Filyanov, Z.\ Phys.\ C {\bf 44}, 157 (1989);
V.~L.~Chernyak and I.~R.~Zhitnitsky, Nucl.\ Phys.\ {\bf B345}, 137
(1990).

\bibitem{BH94} V.~M.~Braun and I.~Halperin, Phys.\ Lett.\ {\bf B328}, 457 (1994).

\bibitem{Br00} V.~M.~Braun, A.~Khodjamirian and M.~Maul, Phys.\ Rev.\ D {\bf 61},
073004 (2000).

\bibitem{Bij} J.~Bijnens and A.~Khodjamirian, Eur.\ Phys.\ J.\ C {\bf 26}, 67 (2002).

\bibitem{Rad80} A.~V.~Efremov and A.~V.~Radyushkin, Phys.\ Lett.\
B {\bf 94}, 245 (1980).

\bibitem{BB99} P.~Ball, V.~M.~Braun, Nucl.\ Phys.\  {\bf B543}, 201 (1999).

\bibitem{Ball99} P.~Ball, J. High Energy Phys. 01 (1999) 010.


\bibitem{An00} J.~R.~Andersen, Phys.\ Lett.\ {\bf B475}, 141 (2000).

\bibitem{BGG} V.~M.~Braun, E.~Gardi and S.~Gottwald, Nucl.\ Phys.\ {\bf B685}, 171
(2004).

\bibitem{Ag051} S.~S.~Agaev, Phys.\ Rev.\ D {\bf 72}, 074020 (2005).

\bibitem{BB1}  P.~Ball, V.~M.~Braun and A.~Lenz, J. High Energy Phys. 05 (2006) 004.


\bibitem{BB2}  P.~Ball, V.~M.~Braun and A.~Lenz, J. High Energy Phys. 08 (2007) 090.

\bibitem{NS94} V.~L.~Chernyak, A.~R.~Zhitnitsky and I.~R.~Zhitnitsky, Yad.\ Fiz.\ {\bf 38}, 1074  (1983) [
Sov.\ J.\ Nucl.\ Phys.\ {\bf 38}, 645 (1983)]; V.~A.~Novikov,
M.~A.~Shifman, N.~G.~Uraltsev and A.~I.~Vainstein, Nucl.\ Phys.\
{\bf B237}, 525 (1984).

\bibitem{Ag052} S.~S.~Agaev, Phys.\ Rev.\ D {\bf 72}, 114010 (2005);
{\bf 73}, 059902(E) (2006); arXiv:hep-ph/0511192.

\bibitem{SVZ} M.~A.~Shifman, A.~I.~Vainstein and V.~I.~Zakharov, Nucl.\
Phys.\ {\bf B147}, 385; 448 (1979).

\bibitem{Beb} C.~J.~Bebek {\it et al.}, Phys.\ Rev.\ D {\bf 17}, 1693 (1978).

\bibitem{Vol} J.~Volmer {\it et al.}, Phys.\ Rev.\ Lett.\ {\bf 86}, 1713 (2001).

\bibitem{Tad} V.~Tadevosyan {\it et al.}, Phys.\ Rev.\ C {\bf 75}, 055205 (2007).

\bibitem{Horn} T.~Horn {\it et al.}, Phys.\ Rev.\ Lett.\ {\bf 97},
192001 (2006).

\bibitem{Agpion} S.~S.~Agaev, Phys.\ Rev.\ D {\bf 69}, 094010
(2004).

\bibitem{choi}H.-M.~Choi and C.-R.~Ji, Phys.\ Rev.\ D {\bf 74},
093010 (2006).

\bibitem{Kh} A.~Khodjamirian, Eur.\ Phys.\ J. C {\bf 6}, 477
(1999).

\bibitem{Sch} A.~Schmedding and O.~Yakovlev, Phys.\ Rev.\ D {\bf
62}, 116002 (2000).

\bibitem{Bak} A.~P.~Bakulev, S.~V.~Mikhailov and N.~G.~Stefanis,
Phys.\ Rev.\ D {\bf 73}, 056002 (2006).

\bibitem{CLEO} J.~Gronberg {\it et al.} (CLEO Collaboration), Phys.\
Rev.\ D {\bf 57}, 33 (1998).






\end{thebibliography}
\end{document}